\documentclass[aps,prl,twocolumn,groupedaddress,showpacs,amsmath,amssymb]{revtex4}
\usepackage{epsfig}
\usepackage{amssymb}
\usepackage{amsmath}
\usepackage{graphicx}
\usepackage{amsfonts}
\usepackage{epsfig}
\usepackage{revsymb}
\usepackage{bm}
\usepackage{latexsym}
\usepackage{hyperref}

\begin{document}

\title{The effect of the interphase kinetics on the motion of a quantum crystal in superfluid liquid}

\author{V. L. Tsymbalenko}
\email[]{vlt49@yandex.ru} \affiliation{National Research Center "Kurchatov Institute", 123182 Moscow, Russia}

%\date{\today}

\begin{abstract}
The fall of a quantum crystal in the state of "burst-like growth" in a superfluid liquid is considered. The experimental data of the pressure variation in the container during the fall of a crystal are discussed. The model of the motion of the crystal is suggested taking the interface dynamics into account.
The results for the numerical simulation of the fall of a crystal are consistent with the experimental data. We find a significant effect of the liquid-crystal  interface kinetics on the hydrodynamics of the liquid flow encircling a crystal.
\end{abstract}


\maketitle

\section{Introduction}
At the temperature $\sim0.1$K the density of the normal component in the superfluid helium is negligible. The motion of a body in superfluid helium at velocities much less than the sound velocity is described by the equations of hydrodynamics of ideal fluid ~\cite{LL}. The specific feature of the  helium crystal motion in the liquid is that the pressure due to the fluid flow produces the chemical potential difference triggering the melting-crystallization processes  at the interface ~\cite{ABP}. The boundary conditions for the hydrodynamic equations, describing the fluid flow,  involve the mass transport across the crystal-liquid interface as well. The shape and size of the crystal vary and the flow pattern of the liquid does as well. This situation differs in kind from the pattern for the motion of a body with an impenetrable surface.

The experiments, performed after the theoretical predictions of the quantum nature of the kinetics at the helium crystal surface  ~\cite{AP}, are focused on the detailed study of the interface kinetics ~\cite{ABP}. Since the crystal is motionless, the stimulation of growth and melting at the interface can  be realized with the various methods. To date, a few experiments are known in which the interface kinetics affects the total crystal surface dynamics. The crystal fixed to the tip at the center of the container and subjected to the hydrostatic pressure gradient melts at the top and crystallizes  at the bottom ~\cite{VLT6}. The picture looks like the crystal moves in the downward direction. The center of the crystal contour shifts downwards at a rate governed by the kinetic growth coefficient.

In paper ~\cite{VLT0} the crystal is placed on the crossbar of the U-shaped superconducting loop. The crystal is either pierced by the crossbar or placed on the platform mounted on the crossbar. The electric current, passed through the superconductor, induces the oscillations of the loop-crystal system. The both methods allow us to achieve the amplitude of the crystal velocity not more than $\sim~3 cm/s$. Above this velocity the crystal pierced by the crossbar starts to melt intensively, then moves downwards and falls to the bottom of the container. In the second method, as the above velocity is reached, the crystal located at the site jumps down to the bottom of the container.

The direct observation of the crystal shape transformation during the fall of the crystal is made by the Japanese group at 0.3 K using a high-speed camera ~\cite{JAP}. The acceleration at which the crystal falls corresponds to the motion of a solid body with the adjoined mass equal to the mass of a half of the liquid displaced by the crystal. A slight deviation takes place at high velocities.

The interest in the motion of a quantum crystal in superfluid helium comes from the nontrivial pressure variation in the container during the crystal fall. The time dependence of the pressure cannot be explained by the movement of the crystal without changing its shape. This means that the phase kinetics at the liquid-crystal interface significantly affects the motion of the falling crystal. In this paper we consider the hydrodynamics of the motion of a quantum crystal with the mobile interface.

\section{Experimental results}
\subsection{Experimental procedure}

\begin{figure}
\begin{center}
\includegraphics[%
  width=0.65\linewidth,
  keepaspectratio]{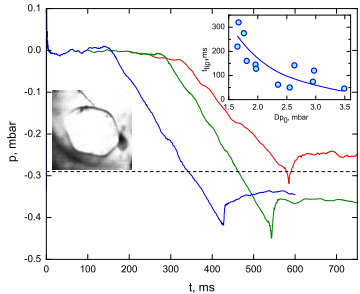}
\end{center}
\caption{The records of the pressure in the container from the nucleation of the crystal to its fall to the bottom of the container. At the initial moment there are pressure oscillations ("burst-like growth"). The horizontal part of the record corresponds to the position of the crystal at the tip. The fall of the crystal is accompanied by some pressure drop. After the crystal reaches the bottom, the pressure relaxes to the equilibrium magnitude (dashed line). The time dependence of locating the crystal at the needle as a function of initial supersaturation is shown in the insert. The image shows the crystal at the moment before the detachment from the needle.}
\label{fig1}
\end{figure}

The nucleation, growth and fall of crystals are observed in an optical dilution refrigerator at  temperature $106$~mK. The crystals are grown in the container with the windows opposite each other at the side walls of the container. The horizontal parallel light beam transmits across the container. The image of the crystal by means of the optical system is focused at the end face of the light guide. The second end face of the light guide is outside the cryostat. The image is scanned with a CCD camera at 25 frames per second. In these experiments the lighting of the crystal is continuous. The end of the tungsten tip lies at the center of the visible field with 12 mm in diameter. This makes it possible to observe the nucleation of a crystal and its detachment from the needle tip.
The pressure in the container is measured with the membrane capacitive sensor located at the upper flange of the container. The parameters of the sensor and the measuring circuit allow one to measure the pressure with the time interval as 40 $\mu$s.

The experiment is performed as follows. First, we set the pressure exceeding the equilibrium one by 0.1-5~mbar. The upper magnitude of pressure is limited with the spontaneous nucleation of crystals at the inner walls of the container. Next, short high-voltage pulse of the $\sim30 \mu$s duration is applied to the needle tip, increasing the local pressure beside the tip due to electrostatic polarization. This entails the nucleation of a crystal at the time moment pointed as t=0 in Fig.1. Depending on the initial supersaturation, the process of crystal growth to the macroscopic size proves to be different. Below the threshold or critical supersaturation the crystal grows sluggishly for $\sim50$~ms (normal growth). Above this supersaturation the pressure drop has an oscillating character and the crystal grows to its maximum size for $\sim0.8$~ms ("abnormal" or "burst-like growth"). The oscillating growth of the crystal decays for the time not longer than 5~ms and is not displayed in the scale of the plot.

The crystal nucleates in the metastable liquid at the tip located in the center of the container, grows to the size of about $1$~mm, hangs on the tip during time $t_{tip}$, then detaches from the support and falls to the bottom. The distance between the tip and the bottom is $17$~mm. This technique is previously used to study the conditions of crystal nucleation in the state of "burst-like growth", see review ~\cite{VLTR}. In the course of the fall the pressure in the container, measured by the sensor on the upper flange, starts to decrease and becomes lower than the difference in the hydrostatic pressure between the center of the container and its bottom of $\Delta p = -0.29$~mbar. Then the pressure increases and reaches the constant value close to the magnitude $\Delta p$, see Fig.1.

In the paper we consider only the crystals grown at the supersaturations above the critical one $\approx0.8$~ mbar and having the high kinetics of crystal facet growth. The kinetic growth coefficient, averaged over the crystal surface, is found from the pressure oscillations. Additional details for the method of determining the kinetic growth coefficient in the oscillating mode are given in review ~\cite{VLTR}, see references there. The magnitudes of growth coefficient are shown in Fig.2. At the end of the oscillating growth the crystal starts melting in the hydrostatic pressure gradient and displaces downwards, resulting finally in its detachment from the needle tip and in the fall to the bottom of the container. The melting time at the needle tip $t_{tip}$ is specified with the position of the kink at the records $p(t)$ with respect to the time moment of crystal nucleation at $t=0$.

\subsection{Growth kinetic coefficient of  a crystal}
At $\sim 0.1$~K the helium crystal has three facet systems, namely: c-facet~$(0001)$, a-facet $(10 \overline 10)$ and s-facet $(10 \overline 11)$. The crystal growth is limited by the growth kinetics at these facets. The phase kinetics will influence the hydrodynamics of the crystal motion provided the kinetic interface growth coefficient $K$ is large. It is important to note that there are two states of the crystal,  which are different in the growth rate of facets. The first state is typical for the crystals with the slow  and normal growth kinetics of facets. The growth rate of the facets for such crystals does not vary in time, increases below $\approx~0.5$K with decreasing the temperature and is non-linear in the supersaturation. The second state of a crystal is abnormal ("burst-like growth"). This state  is a result of the influence of the excessive supersaturation on the crystal within the time interval from milliseconds to tens of seconds. The state is characterized by the high growth rate of all facets, exceeding the growth rate of the facets for the normal crystals by several orders of the magnitude, see review ~\cite{VLTR}. The high growth kinetics of crystal facets remains unchanged for some time after the transition of the crystal to the abnormal state. Then the crystal relaxes into the stable normal state. It is obvious that the influence of the dynamic fluid pressure is most pronounced for the crystals in the anomalous state.

It should be noted that the term "burst-like growth", introduced by the authors of Ref.~\cite{RHBPT}, refers to the crystal c-facet which experiences a jump-like transition from the practically motionless state to the rapid growth state. For the free growth of the helium crystal at the needle, a drastic variation  of the growth kinetics is observed as well. Concerning the crystal nucleated in the metastable liquid, we observe a jump-like enhancement of the growth rate for all facets by 2-3 orders of the magnitude, i.e. «abnormal growth» ~\cite{VLT98}. In Ref.~\cite{VLT99} the general physical mechanism is assumed for these phenomena on the basis of similarity in their manifestations. For the detailed comparison of the effects, see review ~\cite{VLTR}. As for the magnitudes of the kinetic coefficient for the c-facet growth in the "burst-like growth" state, due to limitations of the optical technique and the inertia of the capacitive pressure sensor we only have a lower estimate for the growth coefficient. The basal c-facet, starting from the supersaturation ~100 $\mu$bar,  grows by 200 - 2000 atomic layers for the time less than 1 s, implying $K>10^{-4}$~ s/m, see ~\cite{RHBPT}, p.141. This estimate does not contradict the magnitudes of growth coefficient $K$ obtained for "abnormal growth", see Fig.2. However, this estimate does not allow us to formulate either positive or negative statement about the identity of their physical nature. So, we still continue to use the "burst-like growth" and "abnormal growth" in the same meaning.

The kinetic growth coefficient for the atomically rough surface of a helium crystal is very high at temperature 100 mK. For the surface inclined with 15 degrees from the c-facet, the growth coefficient $K$ is about $5\cdot 10^3$ s/m at 0.3 K. (Ref.~\cite{ABP}, Fig. 48). As the temperature lowers, we have $K\sim 1/T^{3-4}$, that is $K(100mK) > 10^5$ s/m. As is seen from Fig.2, the magnitudes $K$ for the rapid facet growth is much smaller as compared with those inherent in the atomically rough sections. Thus, the shape of the crystal growing at the needle is governed with the slowest sections of  the crystal surface, i.e. with the facets ~\cite{Ch}. By the end of the rapid growth stage the crystal has the shape of a hexagonal prism, see Fig.1. Unfortunately, the continuous lighting, used in the shooting, has not allowed us to obtain the distinct images of crystals in the course of their fall.

In our experiments the crystals are produced in the metastable liquid at high overpressurization when the abnormal state with the high growth kinetics of the facets occurs. By recording the oscillating growth of the crystal immediately after its nucleation, we have determined the kinetic growth coefficient at this time stage, see Fig.2.

It is known from the experiments at high temperatures of $0.48-0.69$K that the crystal returns to the normal slow kinetics of the facets in $\sim 0.1$~s ~\cite{VLT04}. Therefore, the first question to be answered is the following. What is the kinetics of the crystal facets at the time of detaching the crystal from the tip? We estimate the average kinetic growth coefficient during its stay at the tip as follows. The crystal grown on the tip cannot mechanically be shifted down under the influence of gravity. This force is too small but the yield point of the crystal proves to be large ~\cite{VLT77}. The only  way is to melt the crystal in the hydrostatic pressure gradient which leads to the displacement of the crystal down to the breakaway. In Ref.~\cite{VLT6}, the relation between the velocity of the crystal center $v_{center}$, its radius $R$, and the kinetic growth coefficient $K$ reads
 \begin{equation}\label{e00}
    \begin{split}
 v_{center}=-K\frac{\Delta\rho}{\rho'}gR,\;\;  t_{tip}\approx \frac{R}{v_{center}},\;\; \\ K\approx\frac{\rho'}{\Delta\rho}\frac{1}{g t_{tip}}, \;\;
  \Delta \rho = \rho' - \rho\ .
    \end{split}
\end{equation}
Here $\rho$ and $\rho'$ are the densities of the liquid and solid phase, respectively, and $g$ is the acceleration of gravity. The time $t_{tip}$ for the remelting of the crystal and its detachment from the support depends on the position of the crystal at the needle, its shape, and kinetic growth coefficient of the surface. The insert in Fig.1 shows the dependence $t_{tip}$ versus the initial supersaturation at which the crystal is formed. The lower the initial supersaturation, the longer the crystal stays at the needle. Using these data, the kinetic growth coefficient averaged over time $t_{tip}$ is calculated using the formulas ~(\ref{e00}). The results of processing are shown in Fig.2. As can be seen from this graph, the values of the kinetic growth coefficient during remelting are close to those calculated from the pressure oscillations after nucleation at the time t=0

It is seen that the high kinetics of the crystal facets is maintained for 0.1-0.4 s. The coefficient $K$ at the next stage of the process before the breakaway from the tip does not significantly differ  from the magnitudes at the initial stage, see Fig.2. Thus, by the beginning of the fall, the kinetics of the crystal facets remains abnormally fast.

\begin{figure}
\begin{center}
\includegraphics[%
  width=0.65\linewidth,
  keepaspectratio]{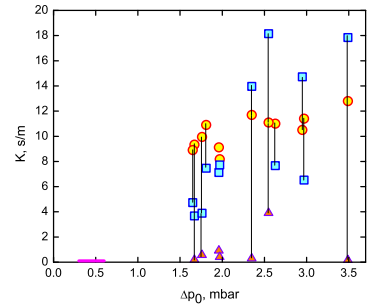}
\end{center}
\caption{The kinetic growth coefficient of crystals. The circles show the values determined by the pressure oscillations, the squares are the values estimated by the time when the crystal slides from the tip, and the triangles are calculated by the pressure relaxation after the stop. The horizontal dashed segment represents the supersaturation interval separating the regions of normal and abnormal growth at $T = 0.1 $K. The vertical lines connect the points that belong to the same crystal.}
\label{fig1}
\end{figure}

As the crystal reaches the bottom of the container, the pressure increases to the equilibrium one, i.e. to the hydrostatic pressure difference $\Delta p$ between the tip and the bottom of the container.
As is shown in Fig.1, the difference between the initial pressure (the crystal at the needle) and the final pressure (the crystal at the bottom) differs from that between the hydrostatic pressures at the needle and at the bottom of the container. The latter is 0.29 mbar within $\pm40 \mu$bar. However, the magnitude 0.29 mbar holds only for the crystals which center before detaching is at the needle tip and which have the same orientation, size and shape after the fall. However, as detaching, the center of a crystal can be located both below and above the needle tip. This changes the height of the fall and, accordingly, the difference of hydrostatic pressure. This difference is also affected with the final crystal orientation and variation in the surface curvature. For example: for the fall of the crystal with the size ratio of 0.5 mm to 2 mm, the displacement of the crystal center lies within 15 -18 mm, entailing uncertainty in the final pressure $\sim 40 \mu$~bar. The variation of the curvature radius of the crystal edges from ~0.25 mm to ~0.2 mm contributes an additional correction $\sim 20 \mu$~bar. These factors may explain the dispersion of the final pressure magnitudes.

Since the relaxation time of the pressure between the container and the external system is of the order of $10$~s, the process observed occurs at almost constant mass of the helium in the container. The pressure growth in the liquid in this case is possible only due to the decrease in the volume of the solid phase. The non-monotonic pressure variation means that the crystal increases its volume during the fall under the influence of the liquid flow. After the stopping the crystal volume returns to the initial one. The growth and melting of the crystal in the container under the fixed helium mass is considered in Ref.~\cite{VLT8}. The pressure relaxation time is determined by the kinetic growth coefficient and is given by the expression
\begin{equation}\label{e01}
\tau_{relax}=\frac{V^{1/3}_{crystal}}{(36 \pi)^{1/3}} \frac{1}{K \Delta p_0}\frac{\rho \rho'}{\Delta\rho}.
\end{equation}
Here $\Delta p_0$ is the supersaturation of the liquid at which the crystal nucleates. The calculation of the kinetic growth coefficient according to  Eq.~(\ref{e01}) is shown in Fig.2. These magnitudes are lower than those obtained before the crystal fall. Unfortunately, it is impossible to draw an unambiguous conclusion about the reason for reducing the kinetic coefficient of crystal facet growth. This can be relaxation similar to that observed at higher temperatures for a fixed crystal (Ref.~\cite{VLT99}). However, we cannot eliminate the influence of the counter flow of a liquid in the process of the fall. In general, the effect can either accelerate or slow down the relaxation of the kinetic growth coefficient to the magnitude typical for the crystals in the normal state.

As we will see from the model calculation, the shape of the crystal varies as well. This results in the relaxation of the crystal shape to the equilibrium one. The relaxation time can be estimated by the eigenfrequency equation of the surface oscillations derived for a spherical quantum crystal ~\cite{BDT,ABS}. The dispersion equation for the oscillation frequencies of the spherical harmonic $L$ reads
 \begin{equation}\label{e02}
 \omega _L^2 - i\omega _L\,\frac{\rho \rho '}{(\Delta \rho)^2 K}\,\frac{L+1}{R}
 -\alpha \frac{ (L^2-1)(L+2)}{R^3}\frac{\rho}{(\Delta\rho) ^2} =0
\end{equation}
where $\alpha$ is the surface tension which we assume to be isotropic for simplicity. The slowest harmonic with $L = 2$ for the values of the kinetic coefficient $K$ within the range from $0.2$ to $5$ s/m (Fig.2) decays exponentially without oscillations at the time constant from 1 to 0.05 s. This value is of the same order of the magnitude as compared with the relaxation time of the crystal volume. The small amplitudes of harmonics do not change the volume of the crystal and do not appear at the  pressure observed.

Thus the crystals, nucleated  in the anomalous state, retain the fast kinetics of crystal facet growth for a long time up to $\sim 0.5$ s. The records of pressure during the crystal fall qualitatively indicate a significant effect of the liquid flow around the crystal on its motion and volume.

\section{Crystal fall in a superfluid liquid}
\subsection{Model approximations}
The shape of the crystal during its growth is determined by the anisotropy of the kinetic growth coefficient $K$ of the crystal facets. The videorecord of the crystal growth in the anomalous state shows that the anisotropy of the growth coefficient $K$ is small. The ratio of its minimum to  maximum magnitudes lies within several units ~\cite{VLTR}. Let us take the growth coefficient $K$ isotropic as is done earlier when processing the data of the oscillating crystal growth in the anomalous state ~\cite{VLT12}. In this approximation the shape of the crystal during its growth under influence of the fluid flow will evolve, tracking only the phase nonequilibrium at the liquid-crystal interface. The initial shape of the crystal is approximated as spherical, this shape being above the roughening transitions ~\cite{VLT6}.

\subsection{The equations of motion of the crystal. The known solution.}
At $\sim 0.1$K the concentration of the normal component of liquid helium is negligible. The motion of the superfluid component at velocities much less than the sound velocity is described by the equations of an ideal fluid ~\cite{LL}. In the fixed reference frame the hydrodynamic equations are given by ~\cite{NU}
\begin{equation}\label{e03}
\overrightarrow v=\nabla \varphi,\;\;    \nabla^2 \varphi = 0,\;\;  p= p_0 -\rho \left(\frac{\partial \varphi}{\partial t} + \frac{1}{2} v^2 +gz\right)\ .
\end{equation}
Here $\overrightarrow v$ is the velocity of the liquid, $\varphi$ is the velocity potential and $z$ is the coordinate axis directed vertically upwards. The boundary conditions read ~\cite{VLT0,NU}
\begin{equation}\label{e04}
\begin{split}
\overrightarrow v=\overrightarrow{u} -\overrightarrow{V}\frac{\Delta \rho}{\rho},\;\; \\
 V=K \delta \mu = K\left[ \frac{\Delta \rho}{\rho' \rho}p - \frac{\alpha}{\rho '}\left (\frac{1}{R_1}+\frac{1}{R_2}\right)+\frac{1}{2}v^2\right]\;\;  .
\end{split}
\end{equation}
Here  $\overrightarrow{u} $ is the velocity of the crystal, $V$ is the normal growth rate of the surface and $R_{1,2}$ are the principal radii of curvature.

In experiments \cite{VLT0} the crystal oscillations occur with the amplitude much smaller than the crystal size. In this case the quadratic terms in  Eqs. (\ref{e03}) and (\ref{e04}) are small. In the linear approximation the crystal shape remains spherical. The force applied to the crystal is given by the expression
\begin{equation}\label{e05}
F = \frac{2 \pi}{3} \rho R^3 \frac{i \omega u}{1+i \omega R \frac{K}{2}\frac{(\Delta \rho) ^2}{\rho \rho'}}.
\end{equation}
One can see from this expression that the phase kinetics at the interface significantly changes the dynamics of the crystal motion. For the small growth kinetic coefficient, the force from the liquid is close to the inertial force of the adjoined mass of the liquid. If growth coefficient $K$ is large, the force is dissipative and vanishes as $K\rightarrow\infty$.

\subsection{The initial phase of the fall.}
After detaching the crystal from the tip, its motion is governed by gravity and buoyancy. The velocity of the crystal and the liquid vanishes. At this stage of the fall the quadratic terms in the system (\ref{e03}) - (\ref{e04}) are small, they can be neglected and the problem becomes linear. Such simplification of the problem is performed under the following conditions:
\begin{equation}\label{e06}
\rho v^2 \ll \rho R a, \;\; a \sim \frac{\Delta \rho}{\rho}g.
\end{equation}
The problem is solved similarly to the problem of the quantum crystal oscillations considered in Ref.~\cite{VLT0}.

The force of gravity, hydrostatic pressure gradient and pressure of the fluid flow pressure exert on the sphere falling in the liquid. The first two factors give the Archimedes law. The resultant force is equal to $\Delta\rho gV_{crystal}$. The flow of the liquid encircles the sphere and the velocity component normal to the sphere surface is given by expression $v_n=2A(t)\cos\theta /R^3$ \cite{LL}. Here $\theta$  is the angle between the $z$ axis and the normal to the surface. Integrating the pressure of the incoming flow and equating it with the Archimedes force, we obtain the equation of motion
\begin{equation}\label{ea01}
\rho'\dot{u} = \Delta\rho - \frac{\rho}{R^3}A(t).
\end{equation}
The boundary condition is derived from relations (5)
\begin{equation}\label{ea02}
\frac{2A(t)}{R^3} =u - \frac{\Delta\rho^2}{\rho\rho'}\frac{K}{R^2}\dot{A}(t).
\end{equation}
Treating these expressions, we arrive at the equations
\begin{equation}\label{ea03}
\dot{A}+\frac{A}{\tau_0} = \frac{\rho}{\Delta\rho}\frac{R^2}{K}gt,\;\; \tau_0 = \frac{\Delta\rho^2}{\rho(2 \rho' +\rho)}KR.
\end{equation}

The crystal starts the motion at zero velocity $u=0$ and motionless liquid $A(0)=0$.
The velocity for the fall of the crystal depends on the time as follows:
\begin{equation}\label{e07}
u(t)=\frac{\Delta \rho}{\rho' + \frac{\rho}{2}}gt + \frac{\rho \Delta \rho}{\rho'(2\rho' + \rho)}g\tau_0 \left[1-exp \left(- \frac{t}{\tau_0}\right)\right].
\end{equation}
The acceleration is equal to
\begin{equation}\label{e08}
a=\dot{u}(t)=\frac{\Delta \rho}{\rho' + \frac{\rho}{2}}g + \frac{\rho \Delta \rho}{\rho'(2\rho' + \rho)} exp \left(- \frac{t}{\tau_0}\right)g.
\end{equation}
The crystal starts to move with the acceleration of gravity, $a(0)=\frac{\Delta\rho}{\rho'}g$. In the experiment for $K>0.2$~s/m, this magnitude is below $1 \mu$s and unobservable in the present measurements. In the steady-state mode $t \gg \tau_0$ the acceleration of the fall coincides with the magnitude of the acceleration of the solid sphere in the ideal liquid if we take the adjoined mass into account ~\cite{LAMB}
\begin{equation}\label{e09}
a_{\infty}=\frac{\Delta \rho}{\rho' + \frac{\rho}{2}}g \approx 0.069g \approx 68\frac{cm}{sec}.
\end{equation}
It is interesting to note that at $K \rightarrow \infty$ the time  of the transition to the steady state goes to infinity and the crystal falls not disturbing the liquid. In this case  the liquid, flowing to the lower side of the crystal, crystallizes without energy dissipation  and  simultaneously the same amount of the solid phase melts from the  opposite side. This conclusion is consistent with the remark to the expression (\ref{e05}) on the disappearance of the force acting on the crystal from the liquid at infinitely high interface kinetics. In the experiment the magnitudes of the kinetic growth coefficient as $\sim 10$ s/m are not so large and such situation is not realized.

\subsection{Next motion of the crystal}
The involvement of quadratic terms into the system of equations (\ref{e03}) - (\ref{e04}) is necessary since condition (\ref{e06}) breaks down. This occurs approximately at $\sim~40$~ms after the start of the fall. Let us consider the situation when the pressure associated with the quadratic terms is much higher than the pressure due to the acceleration of the crystal, $\rho v^2 \gg \rho R a$. The growth rate of the crystal surface is determined by the expression for the difference in the chemical potentials given by Eq.(\ref{e04}). The expression for the pressure reads
\begin{equation}\label{e10}
p=\frac{\rho}{2}(u^2 - v^2) - \frac{\Delta \rho}{\rho}\frac{V_{crystal}}{V_0}k^{-1}_l -\rho gz_{center}.
\end{equation}
Here $V_0$ is the internal volume of the container and $k_l$ is the compressibility of liquid helium. The last two terms in the expression (\ref{e10}) involve: (a)~the pressure drop in the container when the volume of the crystal changes; (b)~the change in the hydrostatic pressure of the liquid when the crystal falls. The qualitative change in the shape of a crystal is clear from the general picture of the liquid flow around the crystal. The flow rate of the liquid is maximum for the cross section which gives the maximum difference in chemical potentials in accordance with formula (\ref{e04}) and, consequently, the maximum growth rate. The crystal starts to increase the transverse size and the volume. As a result, the pressure inside the container decreases faster than it would occur if the crystal falls with the same shape. The growth of the transverse size of the crystal leads to an increase in the adjoined mass of the liquid ~\cite{LAMB} which, in turn, reduces the acceleration at which the crystal falls.

The numerical simulation of the crystal fall, having the constant kinetic growth coefficient $K = 0.22$~s/m, is shown in Fig.3. The shape of the crystal is approximated by an ellipsoid of rotation with the symmetry axis parallel to velocity $u$.
\begin{figure}
\begin{center}
\includegraphics[%
  width=0.65\linewidth,
  keepaspectratio]{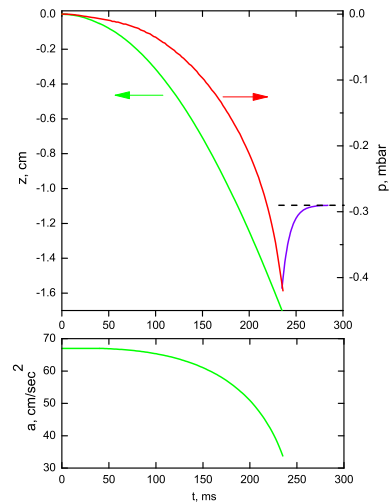}
\end{center}
\caption{The upper graphic: the pressure variation in the container during the crystal fall to the bottom of the container (right scale) and its coordinates with time (left scale). The minimum pressure is lower than the hydrostatic pressure difference between the center and the bottom, which entails an increase in the volume of the crystal during the fall in the fluid flow. The increase in pressure after the stop means the melting of excess volume. The lower graphic is a reduction in the acceleration of the crystal fall due to the change in shape and enhancement of the adjoined mass of the liquid.}
\label{fig1}
\end{figure}
From Fig.3 it can be seen that the total time of the crystal fall to the bottom of the container, distant by $17$ ~mm from the tip at which the crystal nucleates, is about a quarter of second. The time of steady-state acceleration, estimated by formula (\ref{e07}), is about $\tau_0 \sim 1 \mu$~s. The pressure drop is proportional to the displacement of the $z_{center}$ coordinate only in first $\sim 40$~ms, while the crystal retains its shape and volume. The further impact of the counterflow of liquid leads to decreasing the longitudinal size of the crystal and increasing its transverse size. The size ratio at the end of the fall is about 3. Changing the shape of the crystal increases the adjoined mass of the liquid and reduces the acceleration of the crystal fall from $\sim 68$~cm/s$^2$ to $\sim 32$~cm/s$^2$ (see Fig.3, the lower graphic). After the stopping the crystal relaxes to equilibrium and  the excess mass melts. This process is accompanied by an increase in pressure to equilibrium in this position. For $K = 0.22$~s/m, the calculated time is $\tau_{relax}~=~8.1$~ms.

The comparison of the results of our model calculation in Fig.3 with the experimental records in Fig.1 demonstrates that such simplified model of the crystal fall describes the main qualitative features of the process. The phase kinetics plays a significant role even at the magnitudes of the kinetic growth coefficient $K \sim 0.2 s/m$.

The question whether this model is applicable to the interpretation of the experiments of the Japanese group ~\cite{JAP} should be answered negatively. The lower part of the crystal, as is seen in the shooting, has the well-visible facets. The falling crystal is in the normal state with the low kinetics of facet growth. On the contrary, the upper surface of the crystal remains in the atomically rough state. This shape of the crystal differs in kind from the isotropic model accepted.

\section{Conclusion}
A simple model for the fall of a crystal in superfluid liquid with the constant isotropic interface growth kinetic coefficient qualitatively explains the variation in pressure in the container. The numerical estimates, based on this model, allow us to draw a number of significant conclusions on the physical effects accompanying this process. First, it is shown that after the formation of the anomalous state of the helium crystal at temperature $0.1$~K, this state is maintained to a half of second, that is, much longer than at temperatures of $0.48-0.69$~K. Second, the effect of phase kinetics on the motion of a quantum crystal in the ideal fluid is demonstrated. The pressure measurements clearly show an increase in the crystal volume when the liquid flows around the crystal. The change in the shape can only be judged by the results of mathematical modelling. The construction for the quantitative model of the crystal fall in a superfluid liquid requires involving the factors beyond the model considered, such as: the change in the kinetic growth coefficient in time, the crystal faceting which has the specific features in the fluid flow around the crystal edges, and the possible development of instabilities. Accounting for these phenomena is beyond the capabilities of our model.

\begin{acknowledgements}
The author is grateful to S.~T.~Boldarev and R.~B.~Gusev, colleagues at Kapitza Institute for Physical Problems for cooperation in the experiments on the growth of crystals in the optical refrigerator. Special thanks to S.~N.~Burmistrov for fruitful discussions and valuable comments. The author is grateful to V.~S.~Kruglov for interest and support of the work.
\end{acknowledgements}

\pagebreak

\end{document}